\documentclass[aps,pre,floatfix,showpacs,preprint]{revtex4}
\usepackage{graphicx}

\usepackage[greek,english]{babel}
\begin{document}
\latintext
\title{\bf The Quantum-Classical and Mind-Brain Linkages:
        The Quantum Zeno Effect in Binocular Rivalry}
\author{Henry P. Stapp}
\affiliation{Theoretical Physics Group\\
                               Lawrence Berkeley National Laboratory\\
                                         University of California\\
                                     Berkeley, California 94720}
\date{\today}
\begin{abstract}
A quantum  mechanical theory  of the relationship  between perceptions
and brain  dynamics based  on von Neumann's  theory of  measurements is
applied to a recent  quantum theoretical treatment of binocular rivalry
that makes essential use of the  quantum Zeno effect to give good fits
to  the complex available  empirical data.  The often-made  claim that
decoherence  effects in  the  warm, wet,  noisy  brain must  eliminate
quantum effects at the  macroscopic scale pertaining to perceptions is
examined,   and  it   is   argued,  on   the   basis  of   fundamental
principles,  that the  usual decoherence  effects will  not  upset the
quantum   Zeno  effect   that  is   being  exploited   in   the  cited
work.
\end{abstract}
\maketitle
\section{Introduction}
 Efstratios  Manousakis\cite{manousakis}  has   recently  given  a  
quantum  mechanical
 description  of the  phenomena  of binocular  rivalry  that fits  the
 complex empirical data  very well. It rests heavily  upon the quantum
 Zeno effect, which  is a strictly quantum mechanical  effect that has
 elsewhere  been proposed  as the  key feature  that permits  the free
 choices on  the part of  an observer to  influence his or  her bodily
 behavior. The intervention by the observer into the physical dynamics
 is  an essential  feature of  orthodox (Copenhagen  and  von Neumann)
 quantum mechanics.  Within the  von Neumann dynamical  framework this
 intervention  can, with  the  aid  of quantum  Zeno  effect, cause  a
 person's brain  to behave  in a way  that causes  the body to  act in
 accord with the person's  conscious intent. Atmanspacher, Bach, Filk,
 Kornmeier  and Roemer\cite{atman}  have proposed  for the  phenomena  
of bistable
 (Necker  cube) perception  a  theory  that rests  on  an effect  that
 resembles the quantum Zeno  effect. However, their treatment is based
 not  on quantum theory  itself, but  on what  they call  weak quantum
 theory. This is a theory that exhibits a quantum-Zeno-like effect but
 does not  involve Planck's quantum  of action, which is  the quantity
 that characterizes  true quantum effects. The  approach of Manousakis
 would  therefore seem  superior,  because it  uses  the known  actual
 quantum Zeno effect that  arises from orthodox quantum theory itself,
 rather  than upon  a new  conjectural unorthodox  foundation.  On the
 other  hand, using  the  orthodox physics-based  approach might  seem
 problematic,  because   it  depends  on  the  existence   of  a  true
 macroscopic quantum effect  in a warm, wet, noisy,  brain, and it has
 been  argued that  such effects  will be  destroyed  by environmental
 decoherence\cite{decoherence}.  That often cited  argument covers 
many quantum effects,
 but  fails for  fundamental  reasons described in the following  
sections to  upset  the quantum Zeno effect at work here.

\section{Coupled Oscillators in Classical Physics}

It  is   becoming  increasingly   clear  that  our   normal  conscious
experiences are  associated with local  $\sim 40 Hz$ oscillations  of the
electromagnetic fields  at selected  correlated sites on  the cerebral
cortex. These  sites are evidently dynamically coupled,  and the brain
appears to be approximately described by classical physics. So I begin
by  recalling some  elementary  facts about  coupled classical  simple
harmonic oscillators (SHOs).

In suitable units  the Hamiltonian for two SHOs  of the same frequency
is
\begin{eqnarray}
H_0 ={1 \over 2} (p^2_1 + q^2_1 + p^2_2 + q^2_2).
\end{eqnarray}
If we introduce new variables via the canonical transformation
\begin{eqnarray} 
P_1 &=& {1 \over {\sqrt{2}}} (p_1 + q_2)\\
Q_1 &=& {1 \over {\sqrt{2}}}(q_1 - p_2)\\
P_2 &=& {1 \over {\sqrt{2}}} (p_2 + q_1)\\
Q_2 &=& {1 \over {\sqrt{2}}} (q_2 - p_1),
\end{eqnarray}
and replace the above $H_0$ by
\begin{eqnarray}
H = (1+e)( P^2_1 + Q^2_1)/2 + (1-e)( P^2_2 + Q^2_2)/2,
\end{eqnarray}
then this $H$ expressed in the original variables is
\begin{eqnarray}
H = H_0 + e(p_1 q_2 - q_1 p_2).
\end{eqnarray}

If  $e <<1$ then  the term  proportional to  e acts  as a  weak coupling
between the two SHOs whose motions for $e = 0$ would be specified by $H_0$.

The Poisson  bracket (classical) equations  of motion for  the coupled
system are, for any $x$,
\begin{eqnarray}
dx/dt &=& \{ x, H \} = \sum_j \Bigl ( {{\partial x} \over {\partial q_j}} 
{{\partial H} \over {\partial  p_j}} - {{\partial x} \over {\partial p_j}} 
{{\partial H} \over {\partial q_j}}\Bigr ).
\end{eqnarray}
They give
\begin{eqnarray}
dp_1/dt &=& - q_1 + p_2 e,\\
dp_2/dt &=& - q_2 - p_1 e,\\
dq_1/dt &=& p_1 + q_2 e,\\
dq_2/dt &=& p_2 - q_1 e.
\end{eqnarray}
For $e  = 0$ we  have two uncoupled  SHOs, and if  they happen to  be in
phase then we have, for any positive constant $C$, a solution
\begin{eqnarray}
p_1 &=& C \cos (t),\\
q_1 &=& C \sin (t),\\
p_2 &=& C \cos (t),\\
q_1 &=& C \sin (t).
\end{eqnarray}
These equations  specify the  evolving state of  the full system  by a
trajectory  in $(p_1,  q_1, p_2,  q_2 )$  space that,  for each  of  the two
individual systems,  is just a circular  orbit in which  the energy of
that  system  flows  periodically   back  and  forth  between  the  $q^2_i$
coordinate  space (potential  energy) and  $p^2_i$ momentum  space (kinetic
energy) aspects of the system.

For the coupled system, the integration of the time derivatives gives,
up to first order in e and second order in $t$,
\begin{eqnarray}
p_1 &=& C (1 - t^2 /2 + et),\\
p_2 &=& C (1 - t^2 /2 - et),\\
q_1 &=& C (t + et^2/2),\\
q_2 &=& C (t - et^2/2).
\end{eqnarray}

This shows that if  the small coupling $e$ is suddenly turned  on at $t =
0$, then the first-order deviation of the classical trajectory from its
$e = 0$  form will be linear in $Cet$. This  result holds independently of
the relative phase or amplitudes $C > 1$ of the two SHOs.

When  we introduce  the  quantum corrections  by quantizing  this
classical model we obtain  an almost identical quantum mechanical
description  of the  dynamics. In  the  very well  known way  the
Hamiltonian $H_0$ goes over to  (I use units where Planck's constant
is $2\pi$.)
\begin{eqnarray}
  H_0 &=& {1 \over 2} (p^2_1 + q^2_1 + p^2_2 + q^2_2) =
(a^{\dagger}_1 a_1 + 1/2) + (a^{\dagger}_2 a_2 + 1/2).
\end{eqnarray}
The connection  between the classical and quantum  descriptions of the
state of  the system is very  simple: the point  in $(p_1, q_1, p_2, q_2)$
space  that represents  the classical  state  of the  whole system  is
replaced  by  a  ``wave  packet''  that, insofar  as  the  interventions
associated  with  observations can  be  neglected,  is  a smeared  out
(Gaussian) structure centered for all  times exactly on the point that
specifies  the classical  state of  the system.  That is,  the quantum
mechanical  representation  of   the  state  specifies  a  probability
distribution of the  form ($exp(-d^2)$  ) where $d$ is the  distance from a
center (of-the-wave-packet) point $(p_1, q_1,  p_2, q_2$ ), which is, at all
times,  exactly the  point $(p_1,  q_1, p_2,  q_2 )$  that is  the classical
representation of the state.

According to  quantum theory, the operator  $a^{\dagger}_i a_i = N_i$  
is the number
operator that gives  the number of quanta of type i  in the state. The
classical  constant C  appearing  in the  classical  treatment is  the
classical counterpart of $\sqrt{N}$, in  the following sense: if the center of
the wave packet lies at distance C from the origin $(p_1, q_1, p_2, q_2 ) =
(0,  0, 0, 0)$,  then the  ``expectation value''  of $N$  in this  state is
$C^2$. So  the classical and  quantum descriptions are  almost identical:
there is, in the quantum treatment, merely a small smearing-out in $(p,
q)$-space, which is needed to satisfy the uncertainty principle.

This  correspondence  persists  when  the coupling  is  included.  The
coupling term in the Hamiltonian is
\begin{eqnarray}
H_1 &=& e (p_1 q_2 - q_1 p_2 - p_2 q_1 + q_2 p_1)/2 \nonumber \\
   &=& ie/2 (a^{\dagger}_1 a_2 - a_1 a^{\dagger}_2 - a^{\dagger}_2a_1 
+ a_2 a^{\dagger}_1) .
\end{eqnarray}
The  Heisenberg  (commutator) equations  of  motion  generated by  the
quadratic Hamiltonian $H  = H_0+ H_1$ gives the  same equations as before,
but now with operators in  place of numbers. Consequently, the centers
of the wave packets will follow the classical trajectories also in the
$e > 0$ case.

\section{Application}

With these  essentially trivial  calculations out of  the way,  we can
turn to the implied physics.  The above mathematical deductions show a
near identity between the classical  and quantum treatments. If at $t =
0$  we  suddenly  turn  on  the  coupling we  see  that  the  classical
trajectory suddenly departs from the  unperturbed path in a linear (in
time  t) fashion.  In  the classical  case  that is  the full  small-t
story.  But in  orthodox quantum theory there  is, in  principle, an  added
observer-dependent effect.  The observer, in order  to get information
about what  is going  on about him  into his stream  of consciousness,
must initiate  probing actions. According  to the elaboration of the theory
of von Neumann\cite{von-neumann}  described in
Refs\cite{stapp1,stapp2,stapp3,stapp4}, the brain  does  most of  the 
work.  It
creates, in  an essentially mechanical  way, generated by  the quantum
equations of  motion, a proposed query.
Each possible query is  associated with a projection into the
future that  specifies the  brain's computed ``expectation''  about what
its state  will be  after getting the  feedback from the  query (i.e., a
feedback  from  the  associated  act  of  observation.)  The  physical
manifestation of this act is called  ``process 1'' by von Neumann. It is
a  key element  of  the  mathematics associated  with  the process  of
observation:  i.e.,  with  the  entry  into the  observer's
stream of consciousness of information about the state of the physical
world.

In order  to focus on  the key point,  and also to tie  the discussion
comfortably  into   the  understanding  of   neuroscientists  who  are
accustomed to  thinking that the brain  is well described  in terms of
the  concepts  of  classical   physics,  I  shall  consider  first  an
approximation  in  which the  brain  is  well  described by  classical
ideas. Thus the two SHO states  that we are focusing on are considered
to be imbedded in a  classically described brain that is providing the
potential wells  in which these  two SHOs move.  It is the  degrees of
freedom of the brain associated with  these two SHOs that are,  
according to theory
being discussed  here, the neural  correlates of the  consciousness of
the observer  during the  period of the  experiment. Hence it  is they
that are affected by von Neumann's process 1. The remaining degrees of
freedom are treated in  this approximation as providing the background
classically    described    potential    wells    in    which    these
consciousness-related SHOs move. One  of these SHOs corresponds to the
neural correlate of the percept  associated with one eye, the other 
SHO is the neural  correlate of  the percept associated  with the  
other eye. This approach reduces the situation to an exactly solvable 
problem not clouded by the infinity  of effects whose consideration 
usually places any rational  understanding of the  connection between 
mind  and brain beyond our conceptual reach.

In the binocular rivalry context,  let the unperturbed $(e = 0)$ motions
represent the computed  (expected-by-the-brain) evolution of these two
SHO states when  both eyes are seeing essentially  the same scene, and
are  exciting  highly  similar responses,  and  let  the  $e >  0$  case
represent the dynamics of the combined system in the binocular rivalry
case, where  the neural  correlates of the  two possible  percepts are
dissimilar.  The ``expectation''  naturally follows the  unperturbed  
orbit, which
corresponds to normal experience,  in which both eyes view essentially
the  same  scene. But  if  $t  = 0$  represents  the  time  of the  last
observation,  then for  small $t  > 0$  the actual  brain state  in the
rivalry        case   will        diverge        from        the
computed-on-the-basis-of-past-experience state,  due to the 
difference of the actual state from the normal state. 
 In  the binocular rivalry  case,
according to the equations derived  above, the path of the (center of,
and hence the entire) actual Gaussian wave packet will, like its point
classical counterpart, diverge linearly in $t$ from the path expected by
the  brain on  the basis  of past  experience. The  divergence  of the
Gaussian wave packet in the  rivalry case from its ``expected'' circular
orbit is readily visualizable in a two dimensional $(p, q)$ space.

According  to  the  basic  statistical  law  of  quantum  theory,  the
probability that the actual state  of the brain, immediately after the
feedback has occurred, will be in the ``expected'' state is equal to the
square of the  absolute value of the overlap  (integral) of the actual
and  ``expected'' wave  functions.  The collapse  action  occurs in  the
subspace that is associated with the occurring experience. The overlap
of these  two Gaussians is  $(exp(-d^2 /2))$, where $d$ is  the distance
between their centers. Because this distance $d$ increases like $Cet$, the
probability that the actual state will be found at time t to be in the
``expected'' state goes to lowest order  in $t$ like $(1- (Cet)^2/2)$. 
And this result is independent  of the relative phases of  the two oscillators.
The fact  that this probability is  unity minus a  correction of order
$(Cet)$ squared, means that if the probing actions come repetitiously at
time intervals  $\delta t<<1$ such that also  $Ce \delta t <<1$,  
then the probability
that the state will remain on the unperturbed orbit for, say, a second
will remain  high even though the classical  trajectory moves linearly
away from the unperturbed orbit by an  amount of order $Ce$. If 
$\delta t$ is of the order of a few milliseconds,  then the factor 
$Ce$ must be less than
about unity. Because  the number of quanta $N$  in the ``classical'' state
is presumably very large, and $C =  \sqrt{N}$, the coupling e needs to be less
than about  $1/\sqrt{N}$.  

The slowing of  the divergence of  the actual orbit
from the computed (circular-in-this-case)  orbit is a manifestation of
the quantum Zeno effect. The representation in the brain of the posing
of the  question of whether the  state of the neural  correlate of the
occurring  percept is  the  computed/expected state  is von  Neumann's
famous process  1, which  lies at mathematical  core of  von Neumann's
quantum  theory  of  the  relationship between  perception  and  brain
dynamics.

Because  this  argument  is  about  possibilities  that  nature  could
exploit, I  shall consider the  cases where $e  < 1/\sqrt{N}$. 
In  these cases
there will be, in this quasi-classical model, by virtue of the quantum
Zeno effect, a large difference between the observed path predicted by
quantum theory  and the path specified by  the deterministic equations
of classical physics.

I have focused here on the leading powers in t, in order to emphasize,
and exhibit in  a very simple and visualizable way,  the origin of the
key result that for small t on the scale, not of the exceedingly short
period  of the  quantum mechanical  oscillations, nor  even on  $\sim 25 ms$
period of  the $\sim 40 Hz$ scale  of the classical  oscillations, but on
the scale of  the difference of the periods of  the two coupled modes,
there will be, by virtue  of the quantum mechanical effects associated
with the process of rapid repeated observations, a shift from a linear
to a quadratic-in-time  departure of the state of  the system from the
state specified by von Neumann's  process 1. This deviation from unity
corresponds  to   the  factor  $(\cos(et))^2$  in   the  more  complete
probability expression.  Manousakis's work  is based on  these factors
$(\cos(et))^2$ with times $t$ corresponding to intervals between conscious
perceptions of the same scene, and the complementary factor $(\sin(et))^2$
at the termination of repetitions  of the same percept. The success of
Manousakis's  work  is  the  first quantitative  indication  that  von
Neumann's quantum theory of  observation works well in actual practice
at this level of brain dynamics. There is, of course, the powerful
indirect  evidence  stemming,  firstly,  from  the  massive  empirical
successes of orthodox quantum  theory, which uses this collapse theory
of  observation to overcome  the huge  logical difficulties stemming  
from the
uncertainty principle,  and secondly, from  fact that it allows  us to
understand within the framework  of orthodox basic physics the evident
capacity of our conscious  thoughts to influence our physical actions,
and thereby to enter into the process of natural selection.
    
In the  broader quantum mechanical context,  the deterministic quantum
mechanical generalization of the deterministic classical law of motion
generates merely  the set  of possible process  1 actions:  it neither
chooses between the generated  possible process 1 actions, nor selects
the  times t  at  which the  chosen  process 1  actions will  actually
occur. Within the pragmatic  orthodox quantum theory these choices are
therefore treated as, and are called, ``free choices on the part of the
experimenter''. The computations given  above show, in particular, that
the  choices of the  rapidity of  the acts  of observation  can, under
appropriate physical circumstances, and  by virtue of the quantum Zeno
effect, be causally efficacious in the physically described world.

The discussion  has focused  so far  on one very  small region  of the
cortex, or  rather on  one pair of  causally linked regions,  with one
member of  the pair associated  with one possible experience,  and the
other  member   of  the  pair  associated  with   the  rival  possible
experience. But each possible experience is presumably associated with
the   excitation   of   a   large   collection   of   such   localized
regions. Following the principles  of  quantum {\it field} theory the 
quantum state is represented by a tensor product of states associated 
with the
individual  tiny regions.  Each  such region  interacts  with its  own
immediate environment. The mechanism under consideration here does not
involve bringing the ``amplitudes'' located in different tiny regions of
the cortex together, and observing interference effects. Consequently,
the  usual argument\cite{decoherence}  to  the effect  
that  ``decoherence'' effects  will
destroy  quantum effects  has no  immediate bearing  on  the 
{\it basically field theoretic} (as contrasted to ordinary 
first quantized quantum theoretic) situation
being  discussed here.  The  quantum Zeno  effect being  examined here
arises  from the product  ---not the  sum---of the  effects associated
with different local regions. Hence random phase factors attached---by
virtue    of    weak    interactions   with    differing    individual
local environments---to the  above described wave packets associated  
with different regions
do not affect the quantum Zeno effect described here, which are controlled
by the essentially classical trajectories and the absolute values squared of
the overlap integrals.

In the language of the  description used above, the relevant classical
trajectories  will be  in a  space of  a large  number of  doublets of
variables $(p_j, q_j)$ with  many doublets corresponding to cortical sites
associated  with the  image  from  one eye,  and  many other  doublets
corresponding  to cortical  sites associated  with an  image  from the
other  eye.  There  will   be essentially classical couplings  between  
the  trajectories associated  with  one  image  and  the trajectories 
associated  with  the  other image. 
But  the probability considerations  pertaining to the powers  of $t$
that arise from the absolute values of the overlap integrals of the 
Gaussian  wave packets carry 
over directly  to   the  higher-dimensional   case,  due  essentially   to  the
multi-dimensional generalization of  the theorem of Pythagoras. Adding
extra environmentally induced phase factors to the wave packets in the
$(p_j, q_j)$ spaces associated with the different  sites   has  no  
effect  on  the   occurring  product  of probability factors.

In the approximation considered above we have taken into account: (1),
the  effective potential wells  in which  the presumed-to-be-important
correlated  SHO   motions  can  be   considered  to  move;   (2),  the
interactions  with the  environments that  introduce  the uncontrolled
phase shifts that produce the usual environmental decoherence effects;
and (3), the  coupling between the two extended  collective modes that
are  being  excited  by  the  strong  optical  stimuli  from  the  two
eyes.  Within  this  approximation  we  have obtained  a  very  simple
understanding of the  origin and relevance of the  quantum Zeno effect
in  the phenomenon of  binocular rivalry.  Of course,  the brain  is a
complex  system, and  this simple  approximation cannot  be  the whole
story.  But  the  suggestion  here  is  that  this  relatively  simple
quasi-classical   model  displays  the   essence  of   the  quantum 
mechanical understanding of the dynamical connection between (conscious) 
percepts 
and their neural correlates. According  to this approach,  the classical
structure  of  our  experience arises primarily not from environmental 
decoherence effects,  as is  often assumed,  but rather  from the  close 
dynamical connection described above between the quantum and classical 
dynamics of the SHO states that enter into the collapse events that
according to von Neumann's quantum theory  of perception tie our 
experiences to their neural correlates.

This model assumes that the process 1 actions  associated with our
experiences  single  out these pure  quantum states. I have often 
suggested\cite{stapp4,stapp5,stapp6} that the best candidates for the
states  corresponding to the  process  1  actions   associated  with  
our experiences are the so-called coherent states of
of the electromagnetic (Coulomb) field\cite{klauder}.  These states, 
localized  in the array  of excited cortical or other sites corresponding 
to the neural correlates of the occurring thought/percept, are exactly 
what have been  used here. They  are dynamically robust\cite{stapp5},  
and as emphasized above, are closely connected to classically
described  aspects  of  the  brain,  and  thereby  to  the  observer's
classical description of her or his perceptions. The quantum-classical
linkage that is crucial  to the pragmatic success of quantum theory
arises naturally  by connecting brain states to streams of consciousness 
in the way described here.  

The question naturally arises how particular conscious  thoughts come 
to be associated with particular  patterns  of   cortical  excitations.  
Is  some  miracle required  to fix  these  connections? No!  
The  process is  completely natural  and rationally  understandable. 
As  extensively  discussed in Ref.~\cite{stapp1},  and  re-emphasized  
in  Ref.~\cite{stapp2},  trial  and  error
learning   beginning  before  birth   and  involving   feedback  loops
pertaining  to  physically  effective  process  1  actions  that  link
effortful  feelings  to   subsequent  experiences  eventually  produce
conscious awareness of, and  then application of, correlations between
controllable  efforts  and  their  feedbacks, and  these  applications
automatically  strengthen  the  correlations between  intentional  and
perceptual thoughts and the patterns  of brain activity that are their
causal counterparts in the brain. This theory of perception is 
naturalistic and assigns to our conscious thoughts, as components of the 
natural order, a causal role that is closely aligned to the role of our 
thoughts in empirical scientific practice.

\section{Acknowledgements}
This work was supported by  the Director, Office of Science, Office of
High  Energy and  Nuclear Physics,  of the  U.S. Department  of Energy
under contract DE-AC02-05CH11231


\begin{thebibliography}{99}


\bibitem{manousakis}E.   Manousakis, Quantum   theory,  
consciousness and temporal
     perception:  Binocular  rivalry.   Submitted  to  Phys.  Rev.  E.
     [http://arXiv.org/abs/0709.4516]

\bibitem{atman}
H.  Atmanspacher, M.  Bach,  T.  Filk, J.  Kornmeier,  and  H.  Roemer,
Cognitive   time  scales   in   a  Necker-Zeno   model  for   bistable
perception.   To  be   published  in   the  Journal   for  Integrative
Neuroscience. [http://www.igpp.de/English/tda/pdf/neckerzeno.pdf]


\bibitem{decoherence}
H. D. Zeh, On the interpretation of measurement in quantum theory. 
Found. Phys. {\bf 1}, 69-76 (1970),
E. Joos and H. D. Zeh, The emergence of classical properties through interaction
with the environment, Z. Phys. {\bf B5} 223-243 (1985)
D. Giulini, E. Joos, C. Kieffer, J. Kupsch, I.-O Stamatescu, and H. D. Zeh, 
{\it Decoherence and the Appearance of a 
Classical World in Quantum Theory} (Springer, Berlin, Heidelberg, New York, 
1996).
A. J. Leggett, Macroscopic quantum systems and the quantum theory 
of measurement.
Supp. Prog. Theor. Phys. {\bf 69} 80-100 (1980).
M. Tegmark, Importance of quantum decoherence in brain process. 
Phys. Rev. {\bf E 61}, 4194-4206 (2000).


\bibitem{von-neumann} J. Von Neumann, {\it Mathematical Foundations of Quantum 
Mechanics},  Chap. VI, pg. 417 (Princeton University Press, Princeton, 1955).


\bibitem{stapp1}
H. P. Stapp, {\it Mind,  Matter, and Quantum Mechanics}, (Springer, Berlin,
Heidelberg, New York, 1993/2004).


\bibitem{stapp2}
H.   P.  Stapp,  {\it  Mindful  Universe:   Quantum  mechanics   and  the
Participating Observer}, (Springer, Berlin, Heidelberg, New York, 2007).

\bibitem{stapp3}
H.P.   Stapp,  Quantum  interactive   dualism:  An   alternative  to
     materialism,   J.  Consc.   Studies.  12   no.11   43-58  (2005).
     [http://www-physics.lbl.gov/$\sim$stapp/stappfiles.html]

\bibitem{stapp4}
J.  M. Schwartz,  H.P. Stapp, and  M. Beauregard,  Quantum theory  in
neuroscience and  psychology: A neurophysical model  of the mind/brain
interaction. Phil. Trans. Royal Soc. {\bf B 360} (1458) 1306 (2005).

\bibitem{stapp5}
Ref. 3, page 132.

\bibitem{stapp6}
H.P. Stapp,  Light as Foundation of Being,  in {\it Quantum Implications:
Essays in Honor of David Bohm}, (Routledge and Kegan Paul, London and New
York, 1987).

\bibitem{stapp7}
H.P.  Stapp,  On  the  unification  of quantum  theory  and  quantum
physics, in {\it Symposium  on the Foundations of Modern  Physics: 50 years
of the  Einstein-Podolsky-Rosen Gedankenexperiment}, eds.  P. Lahti and
P. Mittelsteadt, (World Scientific, Singapore, 1985).

\bibitem{klauder}
J.R.  Klauder and  E.C.G. Sudarshan,  {\it Fundamentals of  Quantum Optics},
(W.A. Benjamin, New York, 1968).  .
\end{thebibliography}
\end{document}